\pdfoutput=1
\documentclass[12pt,a4paper,titlepage]{article}
\usepackage[utf8]{inputenc}
\usepackage[top=50pt,bottom=50pt,left=68pt,right=66pt]{geometry}
\usepackage{amsmath}
\usepackage{booktabs}
\usepackage{amssymb}
\usepackage[title]{appendix}
\usepackage{mathrsfs}
\usepackage{float}
\usepackage{multirow}
\usepackage{graphicx,caption,subcaption}
\usepackage[space]{grffile}

\begin{document}
\renewcommand{\arraystretch}{1.3}

\makeatletter
\def\@hangfrom#1{\setbox\@tempboxa\hbox{{#1}}%
      \hangindent 0pt
      \noindent\box\@tempboxa}
\makeatother


\def\un#1{\relax\ifmmode\@@underline#1\else
        $\@@underline{\hbox{#1}}$\relax\fi}


\let\under=\unt                 
\let\ced=\ce                    
\let\du=\du                     
\let\um=\Hu                     
\let\sll=\lp                    
\let\Sll=\Lp                    
\let\slo=\os                    
\let\Slo=\Os                    
\let\tie=\ta                    
\let\br=\ub                     


\def\a{\alpha}
\def\b{\beta}
\def\c{\chi}
\def\d{\delta}
\def\e{\epsilon}
\def\f{\phi}
\def\g{\gamma}
\def\h{\eta}
\def\i{\iota}
\def\j{\psi}
\def\k{\kappa}
\def\l{\lambda}
\def\m{\mu}
\def\n{\nu}
\def\o{\omega}
\def\p{\pi}
\def\q{\theta}
\def\r{\rho}
\def\s{\sigma}
\def\t{\tau}
\def\u{\upsilon}
\def\x{\xi}
\def\z{\zeta}
\def\D{\Delta}
\def\F{\Phi}
\def\G{\Gamma}
\def\J{\Psi}
\def\L{\Lambda}
\def\O{\Omega}
\def\P{\Pi}
\def\Q{\Theta}
\def\S{\Sigma}
\def\U{\Upsilon}
\def\X{\Xi}


\def\ve{\varepsilon}
\def\vf{\varphi}
\def\vr{\varrho}
\def\vs{\varsigma}
\def\vq{\vartheta}


\def\ca{{\cal A}}
\def\cb{{\cal B}}
\def\cc{{\cal C}}
\def\cd{{\cal D}}
\def\ce{{\cal E}}
\def\cf{{\cal F}}
\def\cg{{\cal G}}
\def\ch{{\cal H}}
\def\ci{{\cal I}}
\def\cj{{\cal J}}
\def\ck{{\cal K}}
\def\cl{{\cal L}}
\def\cm{{\cal M}}
\def\cn{{\cal N}}
\def\co{{\cal O}}
\def\cp{{\cal P}}
\def\cq{{\cal Q}}
\def\car{{\cal R}}
\def\cs{{\cal S}}
\def\ct{{\cal T}}
\def\cu{{\cal U}}
\def\cv{{\cal V}}
\def\cw{{\cal W}}
\def\cx{{\cal X}}
\def\cy{{\cal Y}}
\def\cz{{\cal Z}}


\def\Sc#1{{\hbox{\sc #1}}}      
\def\Sf#1{{\hbox{\sf #1}}}      



\def\slpa{\slash{\pa}}                            
\def\slin{\SLLash{\in}}                                   
\def\bo{{\raise-.3ex\hbox{\large$\Box$}}}               
\def\cbo{\Sc [}                                         
\def\pa{\partial}                                       
\def\de{\nabla}                                         
\def\dell{\bigtriangledown}                             
\def\su{\sum}                                           
\def\pr{\prod}                                          
\def\iff{\leftrightarrow}                               
\def\conj{{\hbox{\large *}}}                            
\def\ltap{\raisebox{-.4ex}{\rlap{$\sim$}} \raisebox{.4ex}{$<$}}   
\def\gtap{\raisebox{-.4ex}{\rlap{$\sim$}} \raisebox{.4ex}{$>$}}   
\def\TH{{\raise.2ex\hbox{$\displaystyle \bigodot$}\mskip-4.7mu \llap H \;}}
\def\face{{\raise.2ex\hbox{$\displaystyle \bigodot$}\mskip-2.2mu \llap {$\ddot
        \smile$}}}                                      
\def\dg{\sp\dagger}                                     
\def\ddg{\sp\ddagger}                                   

\font\tenex=cmex10 scaled 1200


\def\sp#1{{}^{#1}}                              
\def\sb#1{{}_{#1}}                              
\def\oldsl#1{\rlap/#1}                          
\def\slash#1{\rlap{\hbox{$\mskip 1 mu /$}}#1}      
\def\Slash#1{\rlap{\hbox{$\mskip 3 mu /$}}#1}      
\def\SLash#1{\rlap{\hbox{$\mskip 4.5 mu /$}}#1}    
\def\SLLash#1{\rlap{\hbox{$\mskip 6 mu /$}}#1}      
\def\PMMM#1{\rlap{\hbox{$\mskip 2 mu | $}}#1}   %
\def\PMM#1{\rlap{\hbox{$\mskip 4 mu ~ \mid $}}#1}       %
\def\Tilde#1{\widetilde{#1}}                    
\def\Hat#1{\widehat{#1}}                        
\def\Bar#1{\overline{#1}}                       
\def\sbar#1{\stackrel{*}{\Bar{#1}}}             
\def\bra#1{\left\langle #1\right|}              
\def\ket#1{\left| #1\right\rangle}              
\def\VEV#1{\left\langle #1\right\rangle}        
\def\abs#1{\left| #1\right|}                    
\def\leftrightarrowfill{$\mathsurround=0pt \mathord\leftarrow \mkern-6mu
        \cleaders\hbox{$\mkern-2mu \mathord- \mkern-2mu$}\hfill
        \mkern-6mu \mathord\rightarrow$}
\def\dvec#1{\vbox{\ialign{##\crcr
        \leftrightarrowfill\crcr\noalign{\kern-1pt\nointerlineskip}
        $\hfil\displaystyle{#1}\hfil$\crcr}}}           
\def\dt#1{{\buildrel {\hbox{\LARGE .}} \over {#1}}}     
\def\dtt#1{{\buildrel \bullet \over {#1}}}              
\def\der#1{{\pa \over \pa {#1}}}                
\def\fder#1{{\d \over \d {#1}}}                 


\def\frac#1#2{{\textstyle{#1\over\vphantom2\smash{\raise.20ex
        \hbox{$\scriptstyle{#2}$}}}}}                   
\def\half{\frac12}                                        
\def\sfrac#1#2{{\vphantom1\smash{\lower.5ex\hbox{\small$#1$}}\over
        \vphantom1\smash{\raise.4ex\hbox{\small$#2$}}}} 
\def\bfrac#1#2{{\vphantom1\smash{\lower.5ex\hbox{$#1$}}\over
        \vphantom1\smash{\raise.3ex\hbox{$#2$}}}}       
\def\afrac#1#2{{\vphantom1\smash{\lower.5ex\hbox{$#1$}}\over#2}}    
\def\partder#1#2{{\partial #1\over\partial #2}}   
\def\parvar#1#2{{\d #1\over \d #2}}               
\def\secder#1#2#3{{\partial^2 #1\over\partial #2 \partial #3}}  
\def\on#1#2{\mathop{\null#2}\limits^{#1}}               
\def\bvec#1{\on\leftarrow{#1}}                  
\def\oover#1{\on\circ{#1}}                              

\def\[{\lfloor{\hskip 0.35pt}\!\!\!\lceil}
\def\]{\rfloor{\hskip 0.35pt}\!\!\!\rceil}
\def\Lag{{\cal L}}
\def\du#1#2{_{#1}{}^{#2}}
\def\ud#1#2{^{#1}{}_{#2}}
\def\dud#1#2#3{_{#1}{}^{#2}{}_{#3}}
\def\udu#1#2#3{^{#1}{}_{#2}{}^{#3}}
\def\calD{{\cal D}}
\def\calM{{\cal M}}

\def\szet{{${\scriptstyle \b}$}}
\def\ulA{{\un A}}
\def\ulM{{\underline M}}
\def\cdm{{\Sc D}_{--}}
\def\cdp{{\Sc D}_{++}}
\def\vTheta{\check\Theta}
\def\fracm#1#2{\hbox{\large{${\frac{{#1}}{{#2}}}$}}}
\def\ha{{\fracmm12}}
\def\tr{{\rm tr}}
\def\Tr{{\rm Tr}}
\def\itrema{$\ddot{\scriptstyle 1}$}
\def\ula{{\underline a}} \def\ulb{{\underline b}} \def\ulc{{\underline c}}
\def\uld{{\underline d}} \def\ule{{\underline e}} \def\ulf{{\underline f}}
\def\ulg{{\underline g}}
\def\items#1{\\ \item{[#1]}}
\def\ul{\underline}
\def\un{\underline}
\def\fracmm#1#2{{{#1}\over{#2}}}
\def\footnotew#1{\footnote{\hsize=6.5in {#1}}}
\def\low#1{{\raise -3pt\hbox{${\hskip 0.75pt}\!_{#1}$}}}

\def\Dot#1{\buildrel{_{_{\hskip 0.01in}\bullet}}\over{#1}}
\def\dt#1{\Dot{#1}}

\def\DDot#1{\buildrel{_{_{\hskip 0.01in}\bullet\bullet}}\over{#1}}
\def\ddt#1{\DDot{#1}}

\def\DDDot#1{\buildrel{_{_{\hskip 0.01in}\bullet\bullet\bullet}}\over{#1}}
\def\dddt#1{\DDDot{#1}}

\def\DDDDot#1{\buildrel{_{_{\hskip 
0.01in}\bullet\bullet\bullet\bullet}}\over{#1}}
\def\ddddt#1{\DDDDot{#1}}

\def\Tilde#1{{\widetilde{#1}}\hskip 0.015in}
\def\Hat#1{\widehat{#1}}


\newskip\humongous \humongous=0pt plus 1000pt minus 1000pt
\def\caja{\mathsurround=0pt}
\def\eqalign#1{\,\vcenter{\openup2\jot \caja
        \ialign{\strut \hfil$\displaystyle{##}$&$
        \displaystyle{{}##}$\hfil\crcr#1\crcr}}\,}
\newif\ifdtup
\def\panorama{\global\dtuptrue \openup2\jot \caja
        \everycr{\noalign{\ifdtup \global\dtupfalse
        \vskip-\lineskiplimit \vskip\normallineskiplimit
        \else \penalty\interdisplaylinepenalty \fi}}}
\def\li#1{\panorama \tabskip=\humongous                         
        \halign to\displaywidth{\hfil$\displaystyle{##}$
        \tabskip=0pt&$\displaystyle{{}##}$\hfil
        \tabskip=\humongous&\llap{$##$}\tabskip=0pt
        \crcr#1\crcr}}
\def\eqalignnotwo#1{\panorama \tabskip=\humongous
        \halign to\displaywidth{\hfil$\displaystyle{##}$
        \tabskip=0pt&$\displaystyle{{}##}$
        \tabskip=0pt&$\displaystyle{{}##}$\hfil
        \tabskip=\humongous&\llap{$##$}\tabskip=0pt
        \crcr#1\crcr}}


\def\eV{\,{\rm eV}}
\def\keV{\,{\rm keV}}
\def\MeV{\,{\rm MeV}}
\def\GeV{\,{\rm GeV}}
\def\TeV{\,{\rm TeV}}
\def\sv{\left<\sigma v\right>}
\def\({\left(}
\def\){\right)}
\def\cm{{\,\rm cm}}
\def\K{{\,\rm K}}
\def\kpc{{\,\rm kpc}}
\def\beq{\begin{equation}}
\def\eeq{\end{equation}}
\def\bea{\begin{eqnarray}}
\def\eea{\end{eqnarray}}


\newcommand{\be}{\begin{equation}}
\newcommand{\ee}{\end{equation}}
\newcommand{\nbe}{\begin{equation*}}
\newcommand{\nee}{\end{equation*}}

\newcommand{\fr}{\frac}
\newcommand{\lb}{\label}

\thispagestyle{empty}

{\hbox to\hsize{
\vbox{\noindent July 2022 \hfill IPMU22-0022 \\
\noindent  revised version \hfill }}}

\noindent
\vskip2.0cm
\begin{center}

{\large\bf Formation of primordial black holes after Starobinsky inflation}

\vglue.3in

Daniel Frolovsky~${}^{a}$, Sergei V. Ketov~${}^{a,b,c,d,\#}$ and Sultan Saburov~${}^{a}$
\vglue.3in

${}^a$~Interdisciplinary Research Laboratory, Tomsk State University\\
36 Lenin Avenue, Tomsk 634050, Russia\\
${}^b$~Department of Physics, Tokyo Metropolitan University\\
1-1 Minami-ohsawa, Hachioji-shi, Tokyo 192-0397, Japan \\
${}^c$~Research School of High-Energy Physics, Tomsk Polytechnic University\\
2a Lenin Avenue, Tomsk 634028, Russia\\
${}^d$~Kavli Institute for the Physics and Mathematics of the Universe (WPI)
\\The University of Tokyo Institutes for Advanced Study,  \\ Kashiwa 277-8583, Japan\\
\vglue.1in

${}^{\#}$~ketov@tmu.ac.jp
\end{center}

\vglue.3in

\begin{center}
{\Large\bf Abstract}  
\end{center}

We adapted the Appleby-Battye-Starobinsky model of $F(R)$ gravity towards describing double cosmological inflation and formation of primordial black holes with masses up to $10^{19}$ g in the
single-field model. We found that it is possible to get an enhancement of the power spectrum of scalar curvature perturbations to the level beyond the Hawking (black hole evaporation) limit of $10^{15}$ g, so that the primordial black holes  resulting from gravitational collapse of those large primordial perturbations can survive in the present universe and form part of cold dark matter. Our results agree with the current measurements of cosmic microwave background radiation within $3\s$ but require fine-tuning of the parameters.

\newpage

\section{Introduction}

One of the current candidates for cold dark matter (CDM) is given by primordial black holes (PBH) presumably formed in the early Universe during or after inflation \cite{Novikov:1967tw,Hawking:1971ei,Dolgov:1992pu,Barrow:1992hq,Carr:2003bj,Sasaki:2018dmp,Carr:2020gox,Carr:2020xqk}. 

Cosmological inflation in the early Universe can be described by the Starobinsky model \cite{Starobinsky:1980te}, see also  Ref.~\cite{Ketov:2019toi} for a recent review. The Starobinsky model perfectly fits current observations of the cosmic microwave background (CMB) radiation, relies on gravitational interactions, while its only (inflaton mass) parameter $M$ is fixed by the CMB amplitude. Starobinsky's inflaton (scalaron) has the clear origin as the physical degree of freedom of the higher-derivative gravity, and can be identified with the Nambu-Goldstone boson related to spontaneous breaking of the scale invariance \cite{Ketov:2010qz,Ketov:2012jt}. The current observational constraints on  inflation are available in Refs.~\cite{Planck:2018jri,BICEP:2021xfz,Tristram:2021tvh}.

The Starobinsky model is based on the non-perturbative $(R+R^2/6M^2)$ gravity and does not lead to PBH production, so the question arises whether it can be modified in the context of more general $F(R)$ gravity models in order to generate PBH. The
principal answer to the same question in the context of the dual scalar-tensor (or quintessence) gravity models is known to be in affirmative by demanding the existence of a {\it near-inflection} point in the inflaton scalar potential by using  the {\it double} inflation scenario leading to an enhancement of the power spectrum of scalar perturbations \cite{Garcia-Bellido:2017mdw,Germani:2017bcs,Germani:2018jgr}. In the literature, the single-field models of PBH production are unrelated to the Starobinsky inflation and lead to a reduction of the effective number  of e-folds (relevant to CMB) and the related decrease in the value of the tilt $n_s$ of CMB scalar perturbations, see e.g., 
Refs.~\cite{Garcia-Bellido:2017mdw,Dalianis:2018frf,Ragavendra:2020sop}. For instance, when insisting on the best CMB fit (i.e. the present central observational value) and robust predictions,  one gets the typical PBH masses of the order $10^8$ g \cite{Iacconi:2021ltm} that is well below the Hawking black hole evaporation limit of $10^{15}$ g, so that those light PBH do not survive in the present universe and cannot be part of CDM.

In this letter we revisit those issues in the context of Starobinsky's inflation by relaxing the agreement with the CMB measurements up to $3\sigma$, while simultaneously maximizing the total number of e-folds for double inflation. The double inflation with a near inflection point is achieved by a careful choice of the $F(R)$-gravity function that asymptotically leads to Starobinsky's  $R^2$ term in the large (scalar) curvature limit.

Our paper is organized as follows. In Sec.~2 we introduce our $F(R)$ gravity model and the corresponding scalar potential. Inflationary dynamics is investigated in Sec.~3. In Sec.~4 we study large scalar perturbations and demonstrate the existence of a  significant peak in the power spectrum leading to PBH formation. Sec.~5 is our Conclusion.

\section{The model}

A generic $F(R)$ gravity model leads to singularities in cosmological evolution and does not agree with observations.
A viable cosmological model has to obey the no-ghost or stability conditions, must have a well-defined Newtonian limit, satisfy the Solar system tests, and keep successes of inflation and the standard cosmology. All that makes a choice of new viable $F(R)$ gravity functions non-trivial. 

Those consistency issues were addressed by Appleby, Battye and Starobinsky in Ref.~\cite{Appleby:2009uf} where they proposed a viable $F(R)$ gravity model with
the action
\be \lb{action}
 S = \fracmm{M_{\rm Pl}^2}{2} \int d^4x \sqrt{-g} F(R)~~,
\ee
simultaneously describing the Starobinsky inflation and the present dark energy, by using the following $F$-function:
\begin{equation} \lb{FR}
	F(R)=\epsilon\low{AB} g \ln \left[\fracmm{\cosh \left(\frac{R}{\epsilon_{AB}}-b\right)}{\cosh (b)}\right]
	+(1-g) R+\fracmm{R^2}{6 M^2}~.
\end{equation}
Here $M_{\rm Pl}\sim 10^{18}$ GeV denotes the (reduced) Planck mass, and the $M$ stands for the
inflaton (Starobinsky's scalaron) mass $M\sim 10^{-5} M_{\rm Pl}\sim 10^{13}$ GeV. The $\epsilon\low{AB}$ is defined by
\begin{equation} \lb{epsi}
	\epsilon_\low{AB}=\fracmm{R_0}{2g\ln(1+e^{2b})}~~,
\end{equation}
so that the model has three free parameters: the dimensional parameter $R_0$, and the two dimensionless parameters
$g$ and $b$. In Ref.~\cite{Appleby:2009uf}, the parameter $R_0$ represented the vacuum value of the scalar curvature 
with $\sqrt{R_0}\sim 10^{-33}$ eV, in order to describe the present dark energy, whereas the remaining parameters
$(g,b)$ determined the shape of the inflaton scalar potential. The allowed range of the parameters $(b,g)$ was given in
Fig.~4 of Ref.~\cite{Appleby:2009uf}, which specifies (above the curve) the parameter values needed to get a
metastable de Sitter vacuum.

We use the same function (\ref{FR}) with the important difference in the physical scale and physical interpretation: instead of describing the dark energy we want to  describe double inflation for the purpose of PBH production, by assigning the much higher value of $R_0$ just under the inflationary scale, $\sqrt{R_0}< H_{\rm inf.}\sim 10^{14}$ GeV. The power spectrum is well constrained
on the CMB scale $H_{\rm inf.}$ by observational data, but it is not on smaller scales.

The standard conversion from the $F(R)$ gravity description to the (dual) scalar-tensor gravity description in terms of 
the canonical inflaton scalar field $\phi$ with the potential $V$ is given by~\footnote{See e.g., Ref.~\cite{Ivanov:2021chn} for more details.} 
\begin{equation}\label{duality}
	\phi=\sqrt{\frac{3}{2}}M_{\rm Pl}^2\ln \left(\fracmm{2}{M_{\rm Pl}^2}F'\right)~,\quad  
	V(R)=  M_{\rm Pl}^4 \fracmm{F'R-F}{4(F')^2}~,
	\end{equation}
where the primes denote the derivatives with respect to the argument (scalar curvature $R$).

In the case of the $F$-function (\ref{FR}), we find the potential
\begin{equation} \lb{VR}
	V(R) = \fracmm{y^{2}}{2} \Bigg[\fracmm{R^{2}}{6M^{2}} +gR\tanh \Bigg(\fracmm{R}{\epsilon_{AB}} -b\Bigg) -g\epsilon\low{AB} \ln\Bigg(\fracmm{\cosh(\fracmm{R}{\epsilon_{AB}}-b)}{\cosh(b)}\Bigg)\Bigg]~~,
\end{equation}
where we have used the notation \cite{Ivanov:2021chn} 
\be \lb{ydef}
y\equiv	\exp\left(-\sqrt{\frac{2}{3}}\phi/M_{\rm Pl}\right)  = \Bigg[ 1-g + \fracmm{R}{3M^{2}} +g\tanh \Bigg(\fracmm{R}{\epsilon_{AB}}-b\Bigg)\Bigg]^{-1}~~.
\ee
The quantity $y$ is small during the Starobinsky inflation, so that it can be used as an expansion variable. The scalar
potential $V(\phi)$ cannot be obtained analytically because it requires finding the inverse function $R(\phi)$ from 
Eq.~(\ref{ydef}). However, it is possible numerically.  

\begin{figure}[h]
\begin{minipage}[h]{0.5\linewidth}
\center{\includegraphics[width=0.8\linewidth]{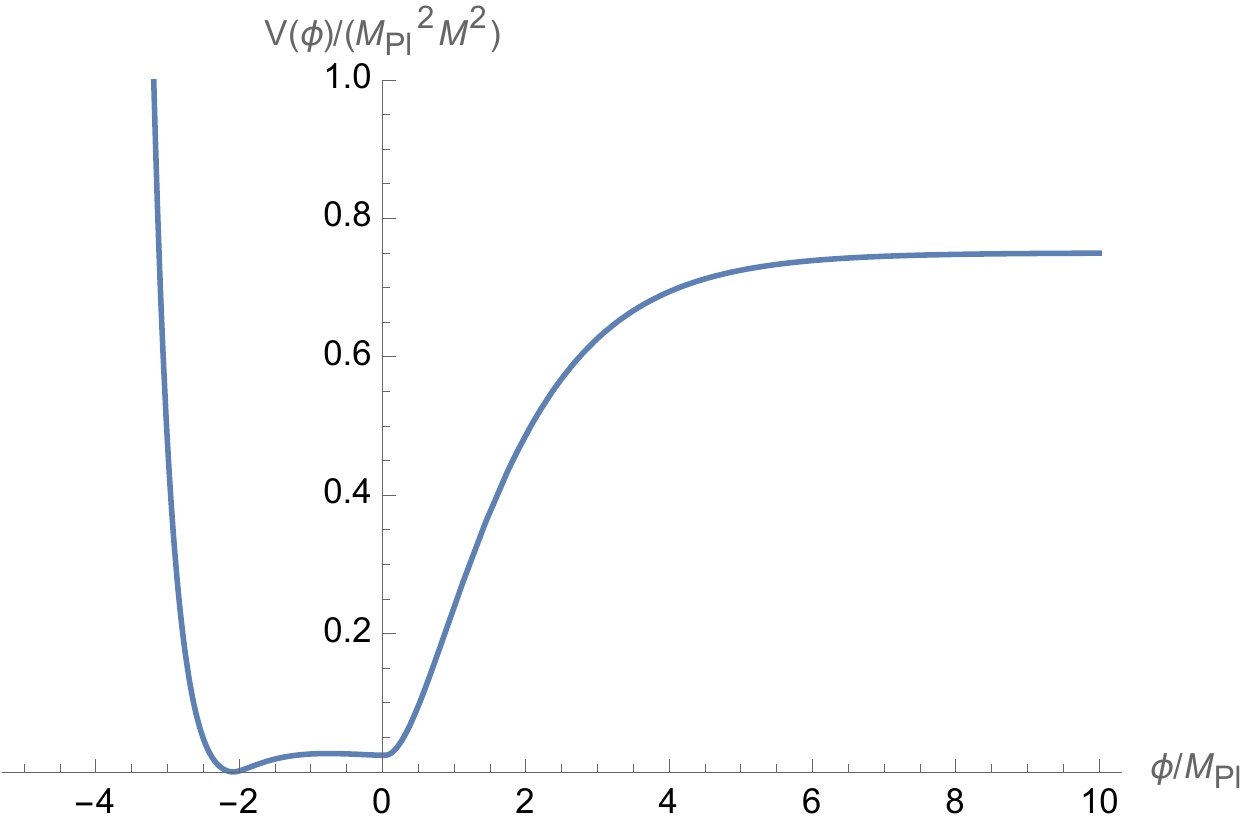} }
\end{minipage}
\hfill
\begin{minipage}[h]{0.5\linewidth}
\center{\includegraphics[width=0.8\linewidth]{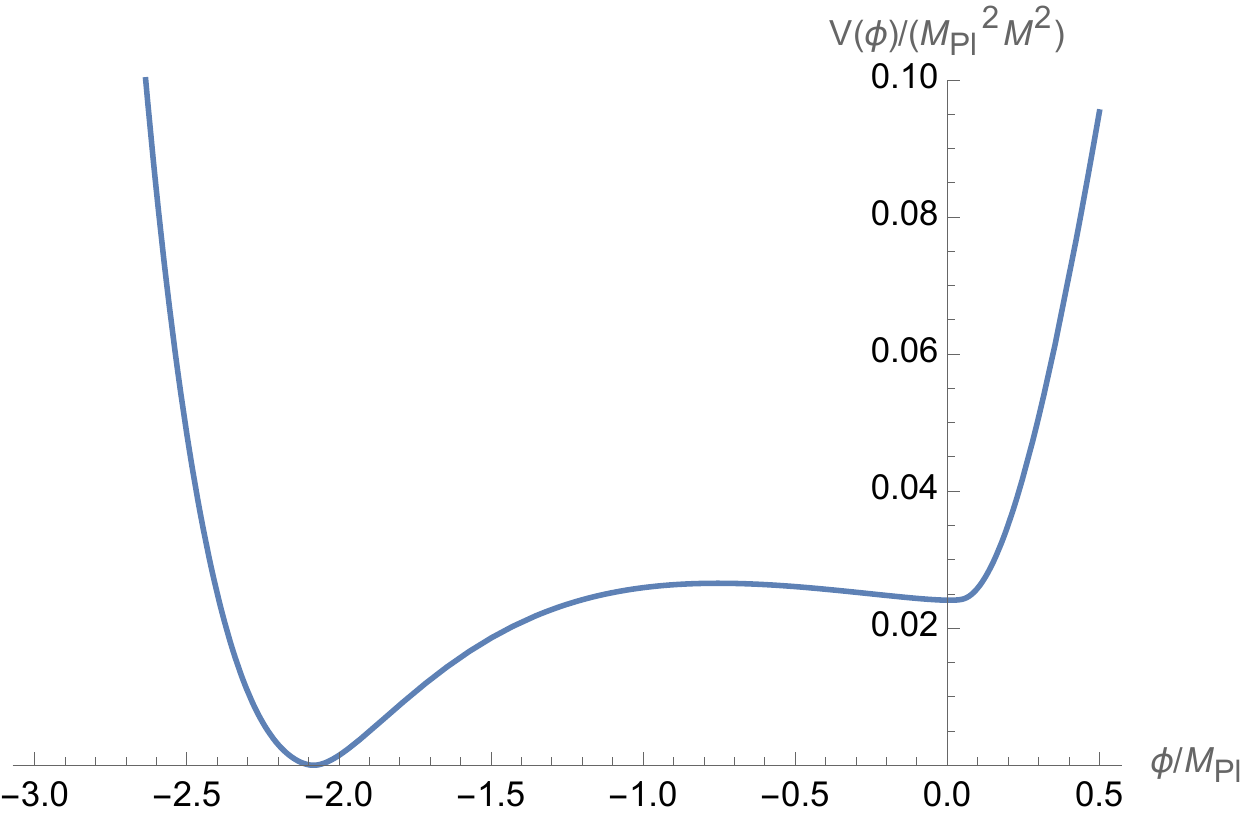} }
\end{minipage}
\caption{ The shape of the inflaton potential $V(\phi)$ (left). Zooming the potential $V(\phi)$ for the selected values
of the inflaton field $\phi$ (right). The parameters are given by $g=0.41$, $b=2.89$  and  $R_0=0.1 M^2$.}
\label{image1}
\end{figure}

The profile of the canonical inflaton potential for the parameter values $g=0.41$, $b=2.89$  and  $R_0=0.1 M^2$ is given in Fig.~1. There are three physically different regimes: (I) the large curvature regime (relevant to CMB) where the potential is the same as that in the Starobinsky model governed by the last term in Eq.~(\ref{FR}), (II) the lower (still high) curvature regime where the scalar potential has a shallow meta-stable de Sitter minimum and a small bump on the left of it, (III) the low curvature regime near the Minkowski vacuum, relevant for reheating after inflation. Actually, the free parameters $(g,b)$ were chosen to get the desired profile of the scalar potential in Fig.~\ref{image1}. 
A significant change in the parameter values may destroy this profile. The regimes (I+II) realize double inflation.

\section{Double inflation}

In this Section we use the standard formulae for describing inflation with a given (canonical) inflaton scalar potential,
see e.g., Ref.~\cite{Ketov:2021fww} for a review. 

The equations of motion are given by
\begin{equation} \lb{eom}
	  \ddot \phi+3H \dot \phi +V'(\phi)=0, \qquad 
	  3H^2=\fracmm{1}{M^2_{\rm Pl}} \left[ \frac{1}{2} \dot \phi ^2+
	   V(\phi) \right]~,\qquad \dot{H} = - \fracmm{1}{2M^2_{\rm Pl}} \dot \phi ^2~~,
	  \end{equation}
where $H(t)$ is Hubble function, and the dots denote the derivatives with respect to time.

The duration of inflation is measured by the e-foldings number
\begin{equation}\lb{efolds}
 	N= \int ^t_{t_{\rm end}} H(t) dt \approx  \fracmm{1}{M_{\rm Pl}^2} \int_{\phi}^{\phi_{in} } \fracmm{V(\phi)}{V'(\phi)} 
	d \phi~~.
 \end{equation}

The slow roll parameters in terms of the scalar potential are given by
\begin{equation} \lb{srV}
	\epsilon(\phi)=\fracmm{M_{\rm Pl}^2}{2}\left[ \fracmm{V'(\phi)}{V(\phi)}\right]^2, \qquad  
	\eta(\phi)=	M_{\rm Pl}^2\left | \fracmm{V''(\phi)}{V(\phi)} \right |~~,
\end{equation}
and they are to be evaluated at the horizon crossing ($\phi_{\rm in}$). We also use the slow-roll parameters in terms of the Hubble flow functions,
\be \lb{srH} \e(t)= - \fracmm{\dot{H}}{H^2}~~, 
\qquad \eta(t)=\fracmm{\dot{\e}}{H\e}~~,
\ee
with the time $t_c$ at the horizon crossing. As the running arguments we use scalar curvature $R$, cosmic time $t$, inflaton field $\phi$ and e-folds $N$ that are all related to each other.

The CMB tilt of scalar perturbations $(n_s)$ and the CMB tensor-to-scalar ratio $(r)$ are given by
\begin{equation} \lb{tilts}
	n_s=1+2\eta(\phi_{\rm in}) -6\epsilon(\phi_{\rm in})= 1- 2\e(t_c) -\eta(t_c)~~, \qquad r=16 \epsilon(\phi_{in})=16 \e(t_c)~.
\end{equation}

\begin{figure}[h]
\begin{minipage}[h]{0.5\linewidth}
\center{\includegraphics[width=0.8\linewidth]{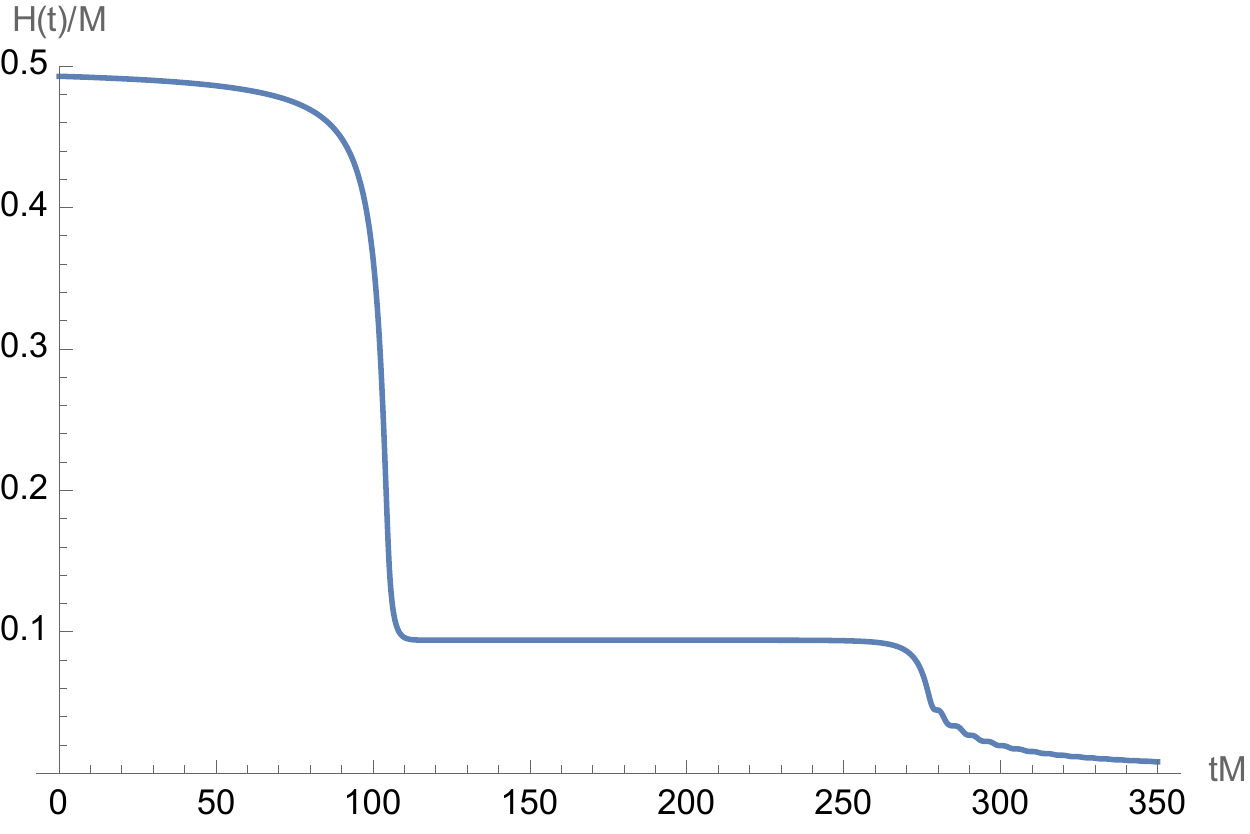} }
\end{minipage}
\hfill
\begin{minipage}[h]{0.5\linewidth}
\center{\includegraphics[width=0.8\linewidth]{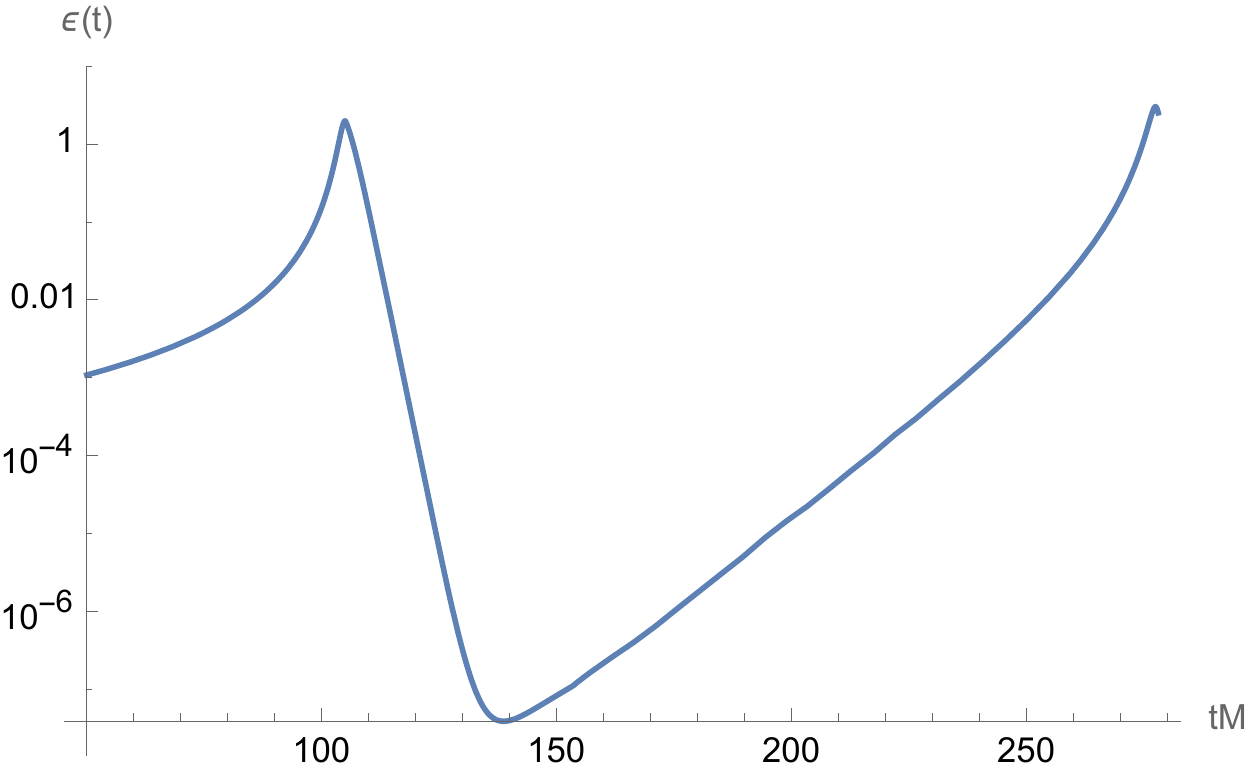} }
\end{minipage}
\caption{On the left: the Hubble function $H(t)$. On the right: the first slow roll parameter $\epsilon(t)$. The initial conditions are $\phi_{\rm in}=5.17 M_{\rm Pl}$  and $\dot\phi_{\rm in}=0$. The end of each stage of inflation is defined by $\epsilon(t_{\rm end})=1$.}
\label{image2}
\end{figure}

Our strategy is to numerically solve the equations of motion by using the initial conditions adapted to the original
Starobinsky model, by choosing  $\phi_{\rm in}\approx 5.17 M_{\rm Pl}$  and $\dot\phi_{\rm in}=0$. Since the Starobinsky 
inflationary solution is an {\it attractor} \cite{Starobinsky:1980te}, the dependence upon local changes of the initial conditions should be irrelevant, except for a duration of inflation. Our numerical solutions to the Hubble function and the first slow roll parameter are given in Fig.~\ref{image2}.

As is clear from Fig.~\ref{image2}, there is the {\it double} inflation indeed, with the two plateaus leading to slow-roll of the inflaton. Between the two stages of slow-roll inflation, there is a short phase of the so-called "ultra-slow-roll" where the first
slow-roll parameter becomes very small, and the slow-roll approximation is broken. Actually, the ultra-slow-roll regime implies fast rolling of the inflaton.

The end of inflation ($\e=1)$ is achieved at $\phi_{\rm end}\approx 0.9 M_{\rm Pl}$  that comes earlier than
$\eta=1$ at $\phi_{\rm end}\approx 0.64 M_{\rm Pl}$. We demand the total e-foldings number, given by a sum of two
slow-roll stages of inflation, $N_{\rm total}=N_e + \Delta N$, to be as large as possible, $N_{\rm total}\approx 60\div 65$.
 
\section{Power spectrum of perturbations and PBH masses}

The simple analytic formula for the power spectrum of scalar (curvature) perturbations 
\cite{Garcia-Bellido:2017mdw}
\begin{equation} \lb{powersp}
	P_{R}=\fracmm{H^2}{8M^2_{\rm Pl}\pi^2\epsilon}
\end{equation}
gives a good approximation as long as the Hubble flow slow-roll parameters (\ref{srH}) are much less
than one. The exact power spectrum should be derived by solving the Mukhanov-Sasaki (MS) equation
\cite{Dalianis:2018frf}. As is clear from our plot on the right side of Fig.~\ref{image2}, Eq.~(\ref{powersp}) does not apply for a small relevant part of the power spectrum with $tM\approx 110\pm 5$.

An ultra-slow-roll regime (also seen on the right side of Fig.~\ref{image2})
leads to an enhancement of the power spectrum, i.e. the appearance of a peak related to large perturbations. The height of the peak should be at least $10^6$ times higher than the reference CMB value $P_R\sim 10^{-9}$, in order to form the PBH surviving until the present times.

\begin{figure}[h]
\center{\includegraphics[width=0.6\linewidth]{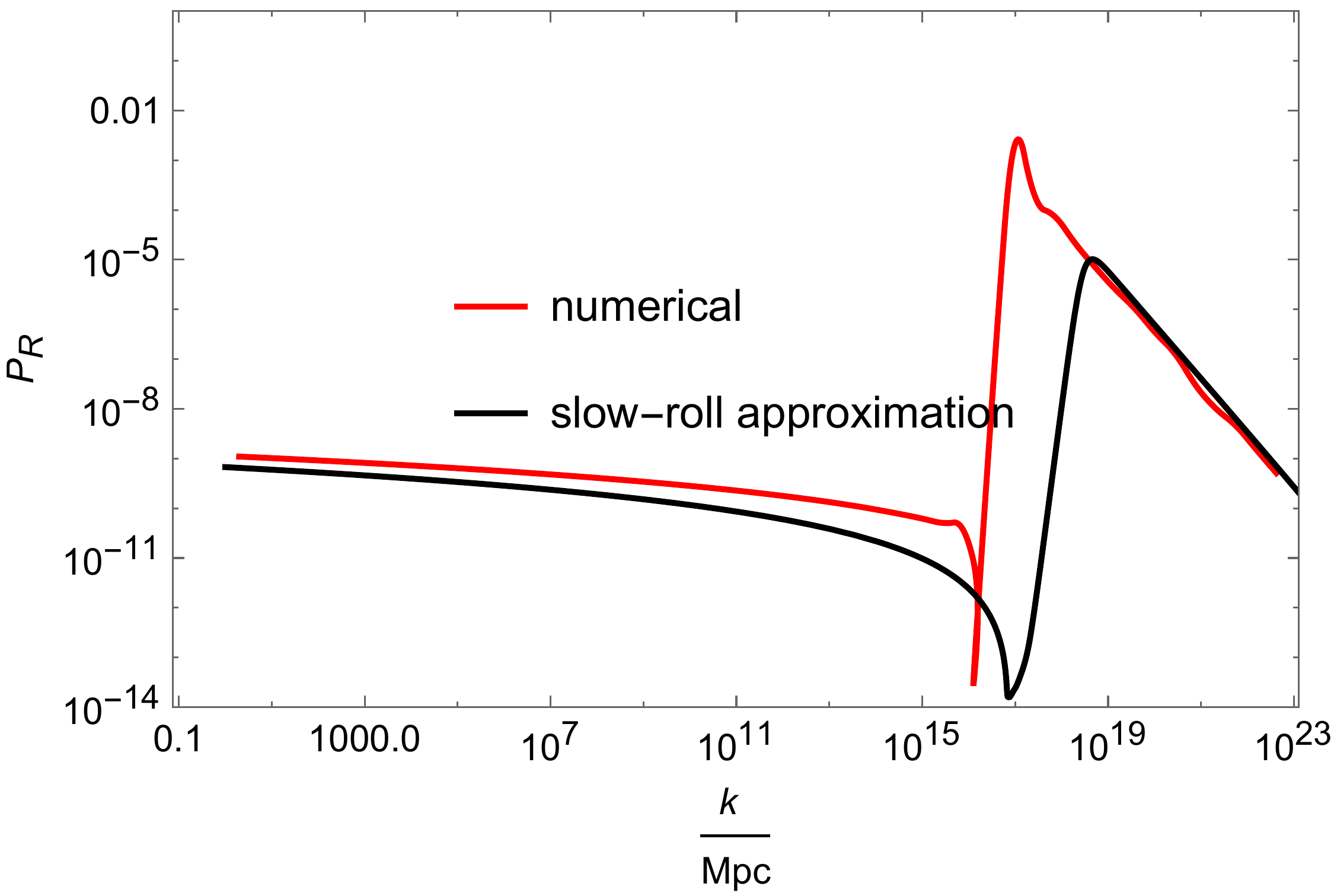}}
\caption{The power spectrum $P_{R}(k)$ of scalar (curvature) perturbations in our model: numerically calculated from the Mukhanov-Sasaki equation (in red) and from Eq.~(\ref{powersp}) in the slow-roll approximation (in black). }
\label{image3}	
\end{figure}

The scalar tilt $n_s$ is related to the power spectrum via the relation $n_s=\fracmm{d \ln P_R}{d\ln k}$ in
terms of the scale variable $k=aH=\dot{a}$. We computed the power spectrum numerically, from 
Eq.~(\ref{powersp}) by using Mathematica and the original code for numerical calculations on a computer. Our results are given by Fig.~\ref{image3}. The height of the peak in Fig.~\ref{image3} corresponds to $P_{R}(k)\sim 10^{-5}$ that is the three orders  of magnitude less than its actual value $P_{R}(k)\sim 10^{-2}$.

The masses of generated PBH can be estimated from the peak as follows~\cite{Pi:2017gih}:
\begin{equation} \lb{pbhm}
M_{\rm PBH}\simeq \fracmm{M_{\rm Pl}^2}{H(t_{\rm peak})} \exp \left[2(N_{\rm total}-N_{\rm peak})+\int_{t_{\rm peak}}^{t_{\rm total} } \epsilon(t) H(t) dt    \right]~~.
\end{equation}
The right-hand-side of this equation is most sensitive to the difference  $(N_{\rm total}-N_{\rm peak})$, while the integral is not very sensitive to the detailed shape of the power spectrum and contributes merely the sub-leading correction. We numerically solved the MS equation, obtained the exact power spectrum and compared it with the result of the slow-roll approximation based on Eq.~(\ref{powersp}), see Fig.~\ref{image3}. The exact (MS-based) peak is the three orders of magnitude higher.

We summarize our findings in the Table where we give the values of the key observables $n_s$, $r$ and $M_{\rm PBH}$ in our model, with the fine-tuned parameters $b=2.89$ and $g=0.41$.
\vglue.2in
\begin{tabular}{ | c | c | c | c | c | c | }
\hline
$\phi_{in}/M_{\rm Pl} $ &$n_s$ & $\Delta N $& $r$ &   $N_{\rm total}$ & $ M_{\rm PBH}$ \\ \hline
  5.099 & 0.95657 & 21 & 0.00532 & 65.03 & $2.17\cdot10^{19}$ g \\
 5.12 & 0.95737 & 20 & 0.005139 & 65.09 & $5.72\cdot10^{18}$ g \\
 5.146 & 0.95831 & 19 & 0.004922 & 65.03 &$ 8.41\cdot10^{17}$ g \\
 5.17 & 0.95915 & 18 & 0.004730 & 65.02 & $1.43\cdot10^{17}$ g \\
  5.095 & 0.95646 & 16 & 0.0053565   &60.01 &$3.89\cdot10^{15}$ g \\
 5.12 &0.95738 &15 & 0.005139   & 60.01 &$7.38\cdot10^{14}$ g \\
\hline
\end{tabular}
\vglue.2in

It follows from the Table that the value of the tensor-to scalar ratio stays well inside the current observational bound, $r<0.036$ ($2\s$). The value of the index $n_s$ of scalar perturbations agrees with the PLANCK measurements~\cite{Planck:2018jri,BICEP:2021xfz,Tristram:2021tvh}, 
\be \lb{planck}
n_s= 0.9649 \pm 0.0042 \quad (1\sigma)~,
\ee
within $2\s\div 3\s$. 

In order to form PBH with masses beyond the Hawking evaporation limit of $10^{15}$ g, so that those PBH can survive in the present universe and form CDM, it is crucial to have the duration $\D N$ of the second stage of inflation for 18 e-folds at least or higher, up to 21. The $M_{\rm PBH}$ grows with increasing $\D N$, but consistency (within $3\s$) with the measured value of $n_s$ is lost beyond $\D N=22$. It is also clear from the Table that the total duration of inflation should be as long as  65 e-folds.

\section{Conclusion}

In this letter we adapted the $F(R)$ gravity model \cite{Appleby:2009uf} for double inflation and PBH production. The significant enhancement of the power spectrum due to large scalar perturbations, namely, the large peak with the height about $10^7$ times larger the CMB reference level is apparently possible. The PBH resulting from gravitational collapse of primordial curvature perturbations can have masses $10^{17}\div 10^{19}$ g, so that they can also survive in the present universe and may form part of CDM. These  PBH masses are very light and smaller than those obtainable in the multi-field models of inflation, see e.g., Ref.~\cite{Ketov:2021fww}.

To get agreement with the CMB measurements within $3\s$, we tuned the parameters $(b,g,R_0)$. The $n_s$ becomes smaller, and $r$ becomes larger, when compared to the original Starobinsky model.

When compared to the results of Ref.~\cite{Garcia-Bellido:2017mdw}, our peak in Fig.~\ref{image3}
is equally high but is narrower. When compared to the results of Ref.~\cite{Iacconi:2021ltm}, our PBH masses (\ref{pbhm}) in the Table are significantly larger. 

It is possible to further generalize the master $F$-function by including a phenomenological description of dark energy, simply by using the Appleby-Battye-Starobinsky ansatz (\ref{FR}) twice, with more parameters $(b_{\rm vac},g_{\rm vac})$ and $\sqrt{R_{\rm vac}}\sim 10^{-33}$ eV.

Fine tuning of the parameters in our model is needed to get a significant enhancement of the power spectrum of scalar perturbations leading to the PBH with masses beyond the Hawking evaporation limit and, hence, the possible PBH DM at present.  A significant change of the parameters would lead to a significant reduction of the power spectrum enhancement with the masses of emerged PBH below the Hawking limit and no PBH DM. However, the remnants of those PBH may still form DM  \cite{Barrow:1992hq}.

Our approach is entirely phenomenological and classical. Considerations of quantum corrections require another framework and are beyond the scope of this investigation (see, however, Refs.~\cite{Liu:2018hno,Ketov:2022lhx}).

\section*{Acknowledgements}

DF and SS were supported by Tomsk State University. DF was also supported by the Foundation for the Advancement of Theoretical Physics and Mathematics "BASIS". SVK was supported by Tokyo Metropolitan University, the Japanese Society for Promotion of Science under the grant No.~22K03624, the World Premier International Research Center Initiative (MEXT, Japan), and the Tomsk Polytechnic University development program Priority-2030-NIP/EB-004-0000-2022. SS was also supported by  the Huawei Technologies.

The authors are grateful to Cristiano Germani, Maxim Khlopov and Ranjan Laha for correspondence.

\bibliography{Bibliography}{}

\providecommand{\href}[2]{#2}\begingroup\raggedright\begin{thebibliography}{10}

\bibitem{Novikov:1967tw}
I.~Novikov and Y.~Zeldovic, ``{Cosmology},''
  \href{http://dx.doi.org/10.1146/annurev.aa.05.090167.003211}{{\em Ann. Rev.
  Astron. Astrophys.} {\bfseries 5} (1967) 627--649}.

\bibitem{Hawking:1971ei}
S.~Hawking, ``{Gravitationally collapsed objects of very low mass},'' {\em Mon.
  Not. Roy. Astron. Soc.} {\bfseries 152} (1971) 75.

\bibitem{Dolgov:1992pu}
A.~Dolgov and J.~Silk, ``{Baryon isocurvature fluctuations at small scales and
  baryonic dark matter},''
  \href{http://dx.doi.org/10.1103/PhysRevD.47.4244}{{\em Phys. Rev. D}
  {\bfseries 47} (1993) 4244--4255}.

\bibitem{Barrow:1992hq}
J.~D. Barrow, E.~J. Copeland, and A.~R. Liddle, ``{The Cosmology of black hole
  relics},'' \href{http://dx.doi.org/10.1103/PhysRevD.46.645}{{\em Phys. Rev.
  D} {\bfseries 46} (1992) 645--657}.

\bibitem{Carr:2003bj}
B.~J. Carr, ``{Primordial black holes as a probe of cosmology and high energy
  physics},'' \href{http://dx.doi.org/10.1007/978-3-540-45230-0\_7}{{\em Lect.
  Notes Phys.} {\bfseries 631} (2003) 301--321},
  \href{http://arxiv.org/abs/astro-ph/0310838}{{\ttfamily
  arXiv:astro-ph/0310838}}.

\bibitem{Sasaki:2018dmp}
M.~Sasaki, T.~Suyama, T.~Tanaka, and S.~Yokoyama, ``{Primordial black
  holes---perspectives in gravitational wave astronomy},''
  \href{http://dx.doi.org/10.1088/1361-6382/aaa7b4}{{\em Class. Quant. Grav.}
  {\bfseries 35} no.~6, (2018) 063001},
  \href{http://arxiv.org/abs/1801.05235}{{\ttfamily arXiv:1801.05235
  [astro-ph.CO]}}.

\bibitem{Carr:2020gox}
B.~Carr, K.~Kohri, Y.~Sendouda, and J.~Yokoyama, ``{Constraints on primordial
  black holes},'' \href{http://dx.doi.org/10.1088/1361-6633/ac1e31}{{\em Rept.
  Prog. Phys.} {\bfseries 84} no.~11, (2021) 116902},
  \href{http://arxiv.org/abs/2002.12778}{{\ttfamily arXiv:2002.12778
  [astro-ph.CO]}}.

\bibitem{Carr:2020xqk}
B.~Carr and F.~Kuhnel, ``{Primordial Black Holes as Dark Matter: Recent
  Developments},''
  \href{http://dx.doi.org/10.1146/annurev-nucl-050520-125911}{{\em Ann. Rev.
  Nucl. Part. Sci.} {\bfseries 70} (2020) 355--394},
  \href{http://arxiv.org/abs/2006.02838}{{\ttfamily arXiv:2006.02838
  [astro-ph.CO]}}.

\bibitem{Starobinsky:1980te}
A.~A. Starobinsky, ``A new type of isotropic cosmological models without
  singularity,''
  \href{http://dx.doi.org/https://doi.org/10.1016/0370-2693(80)90670-X}{{\em
  Phys. Lett. B} {\bfseries 91} no.~1, (1980) 99 -- 102}.

\bibitem{Ketov:2019toi}
S.~V. Ketov, ``{On the equivalence of Starobinsky and Higgs inflationary models
  in gravity and supergravity},''
  \href{http://dx.doi.org/10.1088/1751-8121/ab6a33}{{\em J. Phys. A} {\bfseries
  53} no.~8, (2020) 084001}, \href{http://arxiv.org/abs/1911.01008}{{\ttfamily
  arXiv:1911.01008 [hep-th]}}.

\bibitem{Ketov:2010qz}
S.~V. Ketov and A.~A. Starobinsky, ``{Embedding $(R+R^{2})$-Inflation into
  Supergravity},'' \href{http://dx.doi.org/10.1103/PhysRevD.83.063512}{{\em
  Phys. Rev. D} {\bfseries 83} (2011) 063512},
  \href{http://arxiv.org/abs/1011.0240}{{\ttfamily arXiv:1011.0240 [hep-th]}}.

\bibitem{Ketov:2012jt}
S.~V. Ketov and A.~A. Starobinsky, ``{Inflation and non-minimal
  scalar-curvature coupling in gravity and supergravity},''
  \href{http://dx.doi.org/10.1088/1475-7516/2012/08/022}{{\em JCAP} {\bfseries
  08} (2012) 022}, \href{http://arxiv.org/abs/1203.0805}{{\ttfamily
  arXiv:1203.0805 [hep-th]}}.

\bibitem{Planck:2018jri}
{\bfseries Planck} Collaboration, Y.~Akrami {\em et~al.}, ``{Planck 2018
  results. X. Constraints on inflation},''
  \href{http://dx.doi.org/10.1051/0004-6361/201833887}{{\em Astron. Astrophys.}
  {\bfseries 641} (2020) A10},
  \href{http://arxiv.org/abs/1807.06211}{{\ttfamily arXiv:1807.06211
  [astro-ph.CO]}}.

\bibitem{BICEP:2021xfz}
{\bfseries BICEP, Keck} Collaboration, P.~A.~R. Ade {\em et~al.}, ``{Improved
  Constraints on Primordial Gravitational Waves using Planck, WMAP, and
  BICEP/Keck Observations through the 2018 Observing Season},''
  \href{http://dx.doi.org/10.1103/PhysRevLett.127.151301}{{\em Phys. Rev.
  Lett.} {\bfseries 127} no.~15, (2021) 151301},
  \href{http://arxiv.org/abs/2110.00483}{{\ttfamily arXiv:2110.00483
  [astro-ph.CO]}}.

\bibitem{Tristram:2021tvh}
M.~Tristram {\em et~al.}, ``{Improved limits on the tensor-to-scalar ratio
  using BICEP and Planck},'' \href{http://arxiv.org/abs/2112.07961}{{\ttfamily
  arXiv:2112.07961 [astro-ph.CO]}}.

\bibitem{Garcia-Bellido:2017mdw}
J.~Garcia-Bellido and E.~Ruiz~Morales, ``{Primordial black holes from single
  field models of inflation},''
  \href{http://dx.doi.org/10.1016/j.dark.2017.09.007}{{\em Phys. Dark Univ.}
  {\bfseries 18} (2017) 47--54},
  \href{http://arxiv.org/abs/1702.03901}{{\ttfamily arXiv:1702.03901
  [astro-ph.CO]}}.

\bibitem{Germani:2017bcs}
C.~Germani and T.~Prokopec, ``{On primordial black holes from an inflection
  point},'' \href{http://dx.doi.org/10.1016/j.dark.2017.09.001}{{\em Phys. Dark
  Univ.} {\bfseries 18} (2017) 6--10},
  \href{http://arxiv.org/abs/1706.04226}{{\ttfamily arXiv:1706.04226
  [astro-ph.CO]}}.

\bibitem{Germani:2018jgr}
C.~Germani and I.~Musco, ``{Abundance of Primordial Black Holes Depends on the
  Shape of the Inflationary Power Spectrum},''
  \href{http://dx.doi.org/10.1103/PhysRevLett.122.141302}{{\em Phys. Rev.
  Lett.} {\bfseries 122} no.~14, (2019) 141302},
  \href{http://arxiv.org/abs/1805.04087}{{\ttfamily arXiv:1805.04087
  [astro-ph.CO]}}.

\bibitem{Dalianis:2018frf}
I.~Dalianis, A.~Kehagias, and G.~Tringas, ``{Primordial black holes from
  \ensuremath{\alpha}-attractors},''
  \href{http://dx.doi.org/10.1088/1475-7516/2019/01/037}{{\em JCAP} {\bfseries
  01} (2019) 037}, \href{http://arxiv.org/abs/1805.09483}{{\ttfamily
  arXiv:1805.09483 [astro-ph.CO]}}.

\bibitem{Ragavendra:2020sop}
H.~V. Ragavendra, P.~Saha, L.~Sriramkumar, and J.~Silk, ``{Primordial black
  holes and secondary gravitational waves from ultraslow roll and punctuated
  inflation},'' \href{http://dx.doi.org/10.1103/PhysRevD.103.083510}{{\em Phys.
  Rev. D} {\bfseries 103} no.~8, (2021) 083510},
  \href{http://arxiv.org/abs/2008.12202}{{\ttfamily arXiv:2008.12202
  [astro-ph.CO]}}.

\bibitem{Iacconi:2021ltm}
L.~Iacconi, H.~Assadullahi, M.~Fasiello, and D.~Wands, ``{Revisiting
  small-scale fluctuations in $\alpha$-attractor models of inflation},''
  \href{http://arxiv.org/abs/2112.05092}{{\ttfamily arXiv:2112.05092
  [astro-ph.CO]}}.

\bibitem{Appleby:2009uf}
S.~A. Appleby, R.~A. Battye, and A.~A. Starobinsky, ``{Curing singularities in
  cosmological evolution of F(R) gravity},''
  \href{http://dx.doi.org/10.1088/1475-7516/2010/06/005}{{\em JCAP} {\bfseries
  06} (2010) 005}, \href{http://arxiv.org/abs/0909.1737}{{\ttfamily
  arXiv:0909.1737 [astro-ph.CO]}}.

\bibitem{Ivanov:2021chn}
V.~R. Ivanov, S.~V. Ketov, E.~O. Pozdeeva, and S.~Y. Vernov, ``{Analytic
  extensions of Starobinsky model of inflation},''
  \href{http://dx.doi.org/10.1088/1475-7516/2022/03/058}{{\em JCAP} {\bfseries
  03} no.~03, (2022) 058}, \href{http://arxiv.org/abs/2111.09058}{{\ttfamily
  arXiv:2111.09058 [gr-qc]}}.

\bibitem{Ketov:2021fww}
S.~V. Ketov, ``{Multi-Field versus Single-Field in the Supergravity Models of
  Inflation and Primordial Black Holes},''
  \href{http://dx.doi.org/10.3390/universe7050115}{{\em Universe} {\bfseries 7}
  no.~5, (2021) 115}.

\bibitem{Pi:2017gih}
S.~Pi, Y.-l. Zhang, Q.-G. Huang, and M.~Sasaki, ``{Scalaron from $R^2$-gravity
  as a heavy field},''
  \href{http://dx.doi.org/10.1088/1475-7516/2018/05/042}{{\em JCAP} {\bfseries
  05} (2018) 042}, \href{http://arxiv.org/abs/1712.09896}{{\ttfamily
  arXiv:1712.09896 [astro-ph.CO]}}.

\bibitem{Liu:2018hno}
L.-H. Liu, T.~Prokopec, and A.~A. Starobinsky, ``{Inflation in an effective
  gravitational model and asymptotic safety},''
  \href{http://dx.doi.org/10.1103/PhysRevD.98.043505}{{\em Phys. Rev. D}
  {\bfseries 98} no.~4, (2018) 043505},
  \href{http://arxiv.org/abs/1806.05407}{{\ttfamily arXiv:1806.05407 [gr-qc]}}.

\bibitem{Ketov:2022lhx}
S.~V. Ketov, ``{Starobinsky-Bel-Robinson gravity},''
  \href{http://arxiv.org/abs/2205.13172}{{\ttfamily arXiv:2205.13172 [gr-qc]}}.

\end{thebibliography}\endgroup
\bibliographystyle{utphys}

\end{document}